\documentclass[aps,pre,twocolumn,twoside,floatfix,a4paper,superscriptaddress,showpacs]{revtex4}

\usepackage{bbm}
\usepackage{amsmath, amssymb}
\usepackage{color}
\usepackage{graphicx,epsfig}
\usepackage{dcolumn}
\usepackage{amsmath}
\usepackage{pstricks}
\usepackage{psfrag}
\usepackage{bm}
\usepackage{epsf}

\newcommand{\bra}[1]{\langle #1|}
\newcommand{\ket}[1]{|#1\rangle}
\newcommand{\braket}[2]{\left\langle #1|#2\right\rangle}

\newcommand{\tr}[1]{\mathrm{tr}\left\{#1\right\}}

\newcommand{\la}{\left\langle}
\newcommand{\ra}{\right\rangle}

\newcommand{\td}{\mathrm{d}}

\newcommand{\bla}{bla\\bla\\bla\\bla\\bla}

\newcommand{\PRD}{Phys. Rev. D}
\newcommand{\PRE}{Phys. Rev. E}
\newcommand{\PRL}{Phys. Rev. Lett.}
\newcommand{\EPL}{EPL (Europhys. Lett.)}
\newcommand{\RMP}{Rev. Mod. Phys.}

\newcommand{\mc}[1]{\mathcal{#1}}

\begin{document}

\title{Thermodynamic length for far from equilibrium quantum systems}
\author{Sebastian Deffner}
\affiliation{Department of Chemistry and Biochemistry and Institute for Physical Science and Technology, University of Maryland, 
College Park, Maryland 20742, USA}

\author{Eric Lutz}
\affiliation{Dahlem Center for Complex Quantum Systems, FU Berlin, D-14195 Berlin, Germany}

\date{\today}

\begin{abstract}
We consider a closed quantum system, initially at thermal equilibrium,  driven by arbitrary external parameters. 
We derive a lower bound on the  entropy production which we express in terms of the Bures angle between the nonequilibrium and the corresponding equilibrium state of the system. The Bures angle is an angle between mixed quantum states and defines a thermodynamic length valid arbitrarily far from equilibrium. As an illustration, we treat the case of  a time-dependent harmonic oscillator for which we obtain analytic expressions for generic driving protocols.
 \end{abstract}

\maketitle

Thermodynamics provides a generic framework to describe  properties of  systems at or close to equilibrium. On the other hand, for systems which are far from equilibrium, that is beyond the linear response regime,  no unified formalism has been developed so far. Recently, however,  a number of cold atom experiments have been able to investigate  quantum processes, which occur  far from thermal equilibrium \cite{kin06,hof07,tro12,gri12}; they underline the need for general characterizations of quantum processes that take place beyond the range of linear response theory. In thermodynamics, nonequilibrium phenomena are associated with a non-vanishing entropy production,  $ \la \Sigma\ra=\Delta S-\la Q\ra/T  \geq0$, defined as the difference between the change of entropy   and the (mean) heat divided by temperature \cite{gro84,kon99}.  The positivity of the (mean) entropy production is an expression of the second law of thermodynamics and follows from the Clausius inequality. The entropy production $\la \Sigma\ra$ is expected to be larger the further away from equilibrium a system operates. However, it is not possible to compute $\la \Sigma\ra$, nor  to derive a useful, process-dependent, lower-bound for it, within equilibrium thermodynamics.

For classical systems near equilibrium, such a lower-bound has been obtained using a geometric approach, and expressed in terms of the thermodynamic length \cite{sal83,sal84,nul85}. The latter defines a thermodynamic Riemannian  metric which measures the distinguishability of equilibrium and nonequilibrium distributions \cite{rup95}. In the linear regime, the entropy production is bounded from below by the square of the thermodynamic length  $\la\Sigma\ra\gtrsim\ell^2$. The thermodynamic length plays an important role in finite-time thermodynamics, where it provides limits on the efficiency of thermal machines \cite{and84,and11}. Methods on how to measure $\ell$ have been discussed in Refs.~\cite{cro07,fen09}. Interestingly, the length $\ell$ is identical  to the statistical distance introduced by Wootters  to distinguish two  pure quantum states \cite{woo81}: the angle in Hilbert space between two wave vectors $\psi_1$ and $\psi_2$  is given by  $\ell\left(\psi_1,\psi_2\right) =\arccos{\int \td x\, \sqrt{p_1\left( x\right) \, p_2\left( x\right)}}$, with the two probability distributions $p_1(x) = |\psi_1(x)|^2$ and $p_2(x) = |\psi_2(x)|^2$. Recently, we have extended the notion of thermodynamic length to   closed quantum systems driven arbitrarily far from equilibrium \cite{def10a}.  To this end, we have generalized the length $\ell$  by the Bures angle $\mathcal{L}$  \cite{kak48,bur69,uhl76,joz94,bra94} between the nonequilibrium and the corresponding equilibrium density operators of the system. The Bures metric  is  a generalization of Wootters' metric to mixed quantum states and  plays a major role in quantum information theory \cite{nie00,ben06}. Using the Bures angle,  we have derived  a generalized Clausius inequality with a process-dependent lower bound, $\la \Sigma\ra \geq (8/\pi^2)\, \mathcal{L}^2$, that is valid for arbitrary nonequilibrium driving beyond linear response. This bound, however,  corresponds to the lowest order term of a systematic series expansion as a function of the Bures length.
Our aim in this paper is  to extend our previous findings and derive a sharper  lower bound on the  entropy production $\la \Sigma \ra$ by evaluating the contribution of higher order terms. We then apply this result to the case of a quantum  parametric harmonic oscillator, a model for a driven trapped ion \cite{lei03,hub08}, for which we find exact analytical expressions for the angle $\mathcal{L}$ for arbitrary driving protocol. We furthermore compare these results with those obtained with the trace distance, a non-Riemannian quantum metric \cite{nie00,ben06}. Finally, we derive an upper bound for the quantum entropy production in the Appendix.

\section{Geometric angle between mixed quantum states}

The Bures angle $\mathcal{L}$ is implied by the Bures metric, which formally quantifies the infinitesimal distance between two mixed quantum states described by the density operators $\rho$ and $\rho +d\rho$ as ${\cal L}^2(\rho+\delta\rho,\rho)=\tr{\delta\rho\, G} /2$, where the operator $G$ obeys the equation $\rho\,G+G\,\rho=\delta\rho$ \cite{ben06}.  In the orthonormal basis $\left |i \ra$ that diagonalizes $\rho= \sum_i p_i | i\rangle  \langle i  |$, an explicit expression of the Bures metric is given by ${\cal L}^2(\rho+\delta\rho,\rho)= (1/2) \sum_{i,j}  |\la i|d\rho|j\ra |^2/(p_i+p_j)$. In the limit of pure quantum states, the Bures metric reduces to Wootters' statistical distance, $\ell^2(p,p+dp) = (1/4)\sum_i(dp_i)^2/p_i$ \cite{bra94}.  Wootters' distance is equal to the angle in Hilbert space between two state vectors, and is the only monotone, Riemannian metric (up to a constant factor) which is invariant under {all} unitary transformations \cite{woo81}. It is therefore  a natural metric on the  space of pure states. The Bures metric, on the other hand, being the generalization of Wootters' metric to mixed quantum states  represents a natural, unitarily invariant  Riemannian metric on the space of impure density matrices \cite{ben06}.  

For  any two density operators $\rho_1$ and $\rho_2$, the finite Bures angle $\mathcal{ L}$  is given by
\begin{equation}
\label{e05}
\mathcal{L}\left(\rho_1,\rho_2\right)=\arccos{\left(\sqrt{F\left(\rho_1,\rho_2\right)}\right)}\,,
\end{equation}
where the fidelity $F$ is defined for an arbitrary pair of mixed quantum states as \cite{uhl76,joz94},
\begin{equation}
\label{e06}
F\left( \rho_1,\rho_2\right)=\left[\tr{\sqrt{\sqrt{\rho_1}\,\rho_2\, \sqrt{\rho_1}}} \right]^2.
\end{equation} 
The fidelity is a symmetric, non-negative and unitarily invariant function, which is equal to one only if the two states, $\rho_1$ and $\rho_2$, are identical. For pure quantum states, $\rho_i=\ket{\psi_i}\bra{\psi_i}$, the fidelity reduces to their overlap, $F( \rho_1,\rho_2)=\tr{ \rho_1 \rho_2} = |\braket{\psi_1}{\psi_2} |^2$. It is worth emphasizing that the Bures angle \eqref{e05} is the natural distance quantifying the distinguishability of two density operators. We shall use this property in the following to quantify the distance between a nonequilibrium state and the equilibrium state corresponding to the same configuration of the system. With the help of this thermodynamic length, we will also obtain a lower bound  on the nonequilibrium quantum entropy production.

\section{Thermodynamic length and generalized  Clausius inequality}
We consider a quantum system whose Hamiltonian $H=H_t$ is varied  during a finite time  interval $\tau$. We assume that the system is initially let to equilibrate with a thermal  reservoir at inverse temperature $\beta=1/T$, before an external control parameter is modified. We further assume that the system is quasi-isolated during the finite driving time $\tau$, so that relaxation is negligible and  the dynamics is unitary to an excellent approximation. This corresponds to a realistic experimental situation. For an infinitely large  driving time, much larger than the relaxation time induced by the weak coupling to the reservoir, the transformation is quasistatic and the system remains in an equilibrium state at all times. During such a slow, equilibrium transformation, the  change in free energy,  $\Delta F=\Delta E-T \Delta S$, is equal to the average work $\la W\ra$ done on the quantum system, $ \la W \ra= \Delta F $. Here  $\Delta E=\la H_\tau\ra-\la H_0\ra$ is  the  (internal) energy difference. For a fast, nonequilibrium transformation,  work is larger than the free energy difference. Using the  first law, $\Delta E=\la W\ra+\la Q\ra$,  we can rewrite the nonequilibrium entropy production $ \la \Sigma\ra=\Delta S-\la Q\ra /T$ as,
\begin{equation}
\label{e01}
\la \Sigma\ra=\beta(\la W\ra-\Delta F).
\end{equation}
The  nonequilibrium entropy production $\la \Sigma\ra$ is thus proportional to the difference between the nonequilibrium and the equilibrium work done on the system. Being a mechanical quantity, it is worth noticing that work is always defined, even for far from equilibrium processes.

Let us  denote the density operator of the system at time $t$ by $\rho_t$. The initial equilibrium density operator is then $\rho_0=\exp{\left( -\beta H_0 \right) }/Z_0$, where $Z_0= \tr{ \exp(-\beta H_0)}$ is the initial partition function, with  similar expressions for the equilibrium density operator $\rho_\tau^\text{eq}$ and the partition function $Z_\tau$ at the final time $\tau$. To obtain a microscopic expression for the entropy production, we use $\Delta E = \tr{\rho_\tau H_\tau} - \tr{\rho_0 H_0}$ and note that $-\beta H_{0,\tau}= \ln \rho_{0,\tau} + \ln Z_{0,\tau}$. Combined with the expression $-\beta \Delta F=-\ln{\left(Z_\tau/Z_0\right)}$ for the free energy difference, we find
\begin{equation}
\label{e04}
\la \Sigma\ra=S\left( \rho_\tau||\rho_\tau^\mathrm{eq} \right)=\tr{\rho_\tau\ln{\rho_\tau}-\rho_\tau\ln{\rho_\tau^\mathrm{eq}}},
\end{equation}
where  $S(\rho_\tau||\rho_\tau^\mathrm{eq})$ is the quantum relative entropy \cite{kul78,ume62}. Equation~\eqref{e04} is an exact expression for the nonequilibrium entropy production for  closed quantum systems driven by an external parameter, and a quantum generalization of the classical results presented in Refs.~\cite{kaw07,vai09} (see also Ref.~\cite{def11}).
 We note, however, that the relative entropy is not a true metric, as it is not symmetric and does not satisfy the triangle inequality; it can therefore not be used as a proper quantum distance  \cite{yeu02}.  We next derive a lower bound for the quantum  entropy production which we express in terms of the Bures angle \eqref{e05}.

Inequalities are important tools of classical and quantum information theory,  as  they allow to express 'impossibilities', that is things that cannot happen \cite{yeu02}. An elementary example is Klein's inequality,  $S( \rho_1||\rho_2) \geq 0$, which expresses the non-negativity of  the quantum relative entropy \cite{nie00}. Combined with  Eq.~\eqref{e04}, it immediately leads to the usual Clausius inequality. A generalized Clausius inequality can be derived by noting that the quantum relative entropy satisfies (Ref.~\cite{aud05}, Theorem 4), 
\begin{equation}
\label{e07}
S(\rho_1||\rho_2)\geq s\left(\frac{d\left(\rho_1,\rho_2\right)}{d\left(e^{1,1},e^{2,2}\right)}\right)\,,
\end{equation}
 if $d(\rho_1,\rho_2)$ is an unitarily invariant  norm. Further, $e^{i,j}=\ket{i}\bra{j}$ is the matrix with $i, j$ elements equal to 1 and all other elements 0. The lower bound \eqref{e07} has been derived with the help of optimization theory and is therefore  as sharp as possible. The function $s(x)$ is explicitly given by the expression \cite{aud05},
\begin{equation}
\label{e08}
\begin{split}
s(x)&=\min\limits_{x<r<1}\Big\{\left(1 - r + x\right) \log{\left(1+\frac{x}{1-r}\right)}
\\&+ \left(r - x\right) \log{\left(1 - \frac{x}{r}\right)}\Big\}.
\end{split}
\end{equation}
The first five nonzero terms in a series expansion around the origin $x=0$ read,
\begin{equation}
\label{e09}
s(x)=2 x^2 +\frac{4}{9}x^4+\frac{32}{135}x^6+\frac{992}{5103}x^8+\frac{6656}{32805}x^{10}+O(x^{12})\,.
\end{equation}
Applying inequality \eqref{e07} to the unitarily invariant Bures angle $\mathcal{L}$, we obtain a process-dependent lower bound on the nonequilibrium entropy production. Taking into account  that  $\mathcal{L}(e^{1,1},e^{2,2})=\pi/2$, since the two matrices $e^{1,1}$ and $e^{2,2}$ are orthogonal $[F(e^{1,1},e^{2,2})=0]$, we find,
\begin{equation}
\label{e10}
\la \Sigma\ra \geq s\left(\frac{2}{\pi} \,\mathcal{L}\left(\rho_\tau,\rho_\tau^\mathrm{eq}\right)\right)\geq\frac{8}{\pi^2} \,\mathcal{L}^2\left(\rho_\tau,\rho_\tau^\mathrm{eq}\right).
\end{equation}
The first order term in the expansion \eqref{e09} yields the generalized Clausius inequality, $\la \Sigma\ra \geq (8/\pi^2)\, \mathcal{L}^2$, derived in Ref.~\cite{def10a}. Since  the terms in the expansion \eqref{e09} are positive, an increasingly sharper lower bound can be obtained by taking more terms into account \cite{rem1}.  An illustration for the case of a  quantum harmonic oscillator with time-dependent frequency, to be discussed in detail in the next Section, is shown in Fig.~1.

Equation \eqref{e10} indicates that the  nonequilibrium  entropy production $\la \Sigma\ra$ is bounded from below by a function of the geometric distance  between the actual density operator $\rho_\tau$  of the system at the end of the driving  and the corresponding equilibrium operator $\rho_\tau^\mathrm{eq}$, as measured by the Bures angle.  Thus the Bures angle provides a natural scale to compare $\la \Sigma\ra$ with, and quantifies in a precise manner the  notion that the  entropy production is larger when a system is driven farther away from equilibrium. 

In the classical limit, where nonequilibrium and equilibrium states are diagonal in the energy  basis, the Bures angle reduces to Wootters' statistical distance. As a result,  Eq.~\eqref{e10} yields a lower bound to the classical nonequilibrium entropy production that is valid for any nonequilibrium driving beyond the linear response regime,
\begin{equation}
\label{e10a}
\la \Sigma\ra_\text{cl} \geq s\left(\frac{1}{2\pi} \,\ell \left(p_\tau,p_\tau^\mathrm{eq}\right)\right)\geq\frac{2}{\pi^2}\, \ell^2\left(p_\tau,p_\tau^\mathrm{eq}\right).
\end{equation}
Moreover, when nonequilibrium and equilibrium states are infinitesimally close, Eq.~\eqref{e10a} takes the form $\la \Sigma\ra_\text{cl} \geq (2/\pi^2) d\ell^2 $, which has been obtained in Refs.~\cite{sal83,sal84,nul85}. It is worth emphasizing that the latter was derived by  expanding the entropy around equilibrium to second order; it is therefore only valid  in the linear response regime.

\section{Parametric harmonic oscillator}

Let us now apply the generalized Clausius inequality \eqref{e10} to the case of a time-dependent harmonic oscillator. The latter provides an important physical model for many quantum systems, for example ultracold trapped ions \cite{lei03,hub08}, and is furthermore analytically solvable. We will, in particular, evaluate the Bures angle \eqref{e05} for a non-trivial quantum time evolution. 
Explicit expressions for $\mathcal{L}$ are in general only known for low dimensional systems \cite{hub93,sla96,dit99}. The difficulty arises from the operational square roots in the definition of the quantum fidelity \eqref{e06}. For Gaussian states, however, the expression for the fidelity simplifies  and can be written in closed form \cite{scu98}.

 The Hamiltonian of the parametric quantum harmonic oscillator is of the usual  form ($M$ denotes the mass),
\begin{equation}
\label{e18}
H_t=\frac{p^2}{2M}+\frac{M}{2}\omega^2_t x^2.
\end{equation}
We assume that the  time-dependent frequency $\omega_t$ starts with initial value $\omega_0$ at $t=0$ and ends with  final value $\omega_1$ at $t=\tau$.  Due to the quadratic form of the Hamiltonian \eqref{e18}, the wavefunction of the  oscillator is Gaussian for any driving protocol $\omega_t$. By introducing the  Gaussian wave function ansatz \cite{hus53},
\begin{equation}
\label{e19}
\psi_t(x)=\exp{\left[\frac{i}{2\hbar}\left(a_t x^2+2b_t x+c_t\right)\right]}, 
\end{equation}
the  Schr\"odinger equation for the quantum oscillator can be reduced to  a system of three coupled differential equations for the time-dependent coefficients $a_t$, $b_t$ and $c_t$,
\begin{subequations}
\begin{eqnarray}
\label{e20a}
\frac{1}{M}\,\td_t a_t&=&-\frac{a^2_t}{M^2}-\omega^2_t,\\
\label{e20b}
\td_t b_t&=&-\frac{a_t}{M}\,b_t  ,\\ 
\label{e20c}
\td_t c_t&=&i\hbar\frac{a_t}{M}-\frac{1}{M}\,b^2_t  .
\end{eqnarray}
\end{subequations}
The nonlinear equation \eqref{e20a} is of the Riccati type. It can be mapped onto the equation of motion of a classical, {force free},  time-dependent harmonic oscillator via the transformation $a_t=M\dot X_t/X_t$. The resulting equation reads $\ddot X_t+\omega^2_t\,X_t=0$. Equations \eqref{e20b}-\eqref{e20c} can be solved once the solution of Eq.~\eqref{e20a} has been determined. With the solutions of the three equations \eqref{e20a}-\eqref{e20c} known, the Gaussian wave function  $\psi_t(x)$ \eqref{e19} is fully characterized by the time-dependence of the angular frequency $\omega_t$. It can be shown  that the dynamics  is  completely determined  by the function $Q^{*}$ introduced by Husimi \cite{,hus53,def08,def10}, 
\begin{equation}
\label{e21}
Q^{*}=\frac{1}{2\omega_0\omega_1}\left[\omega_0^2\,\left(\omega_1^2\,X^2_\tau+\dot X^2_\tau\right)+\left(\omega_1^2\,Y^2_\tau+\dot Y^2_\tau\right)\right]\,, 
\end{equation}
where $X_t$ and $Y_t$ are the solutions of the force free classical oscillator  equation satisfying the boundary conditions $X_0=0$, $\dot X_0=1$ and $Y_0=1$, $\dot Y_0=0$. The function $Q^*\geq1$ is a measure of the adiabaticy of the process: it is  equal to one for  adiabatic transformations and increases with the degree of nonadiabaticty.  In particular, the final mean energy of the quantum oscillator is given by \cite{def10},
\begin{equation}
\label{e22}
\la H_\tau\ra=\frac{\hbar\omega_1}{2}\,Q^*\coth(\beta\hbar\omega_0/2),
\end{equation}
and thus linearly increases with $Q^*$.

To evaluate the Bures angle \eqref{e05} for the parametric harmonic oscillator, the quantum fidelity  \eqref{e06} has to be written in  closed form. For Gaussian states such an explicit form is known: for  two arbitrary (non-displaced)  Gaussian density operators, $\rho_1$ and $\rho_2$, the fidelity reads \cite{scu98},
\begin{equation}
\label{e23}
F\left(\rho_1,\rho_2\right)=\frac{2}{\sqrt{\Delta+\delta}-\sqrt{\delta}}\,.
\end{equation}
The two parameters $\Delta=\det{\left(A_1+A_2\right)}$ and $\delta=\left(\det{\left(A_1\right)}-1\right)\left(\det{\left(A_2\right)}-1 \right)$ are completely determined by the  covariance matrices $A_i$ (matrices of the variances of position and momentum) of the quantum oscillator,
\begin{equation}
\label{e24}
A_i=\begin{pmatrix}a_{xx}^i&a_{xp}^i\\
a_{xp}^i&a_{pp}^i
\end{pmatrix}\,.
\end{equation}
The matrix elements $a^i$ are explicitly given by
\begin{subequations}
\label{e25}
\begin{eqnarray}
\label{e25a}
a_{xx}^i&=&2\left(\la x_i^2\ra-\la x_i\ra^2 \right) \\
\label{e25b}
a_{pp}^i&=& \frac{2}{\hbar^2}\left(\la p_i^2\ra-\la p_i\ra^2 \right)\\
\label{e25c}
a_{xp}^i&=&\frac{2}{\hbar}\left(\frac{1}{2}\la x_i p_i+p_i x_i\ra-\la x_i\ra\la p_i\ra \right) \,.
\end{eqnarray}
\end{subequations}
To evaluate   the terms appearing in the Clausius inequality \eqref{e10}, we make use of the explicit expressions of the initial, $\rho_0$, and final density operators, $\rho_\tau$ and $\rho_\tau^\mathrm{eq}$, of the oscillator in coordinate representation, as  given in Appendix C. In particular, the final equilibrium density operator $\rho_\tau^\mathrm{eq}$ has  the same form  as  Eq.~\eqref{e40}, replacing $\omega_0$ by $\omega_1$. Accordingly,  the corresponding equilibrium mean and variances are $\la x\ra_{\tau}^\mathrm{eq}=\la p\ra_{\tau}^\mathrm{eq}=\la xp+px \ra_\tau^\mathrm{eq}=0$, and 
\begin{subequations}
\begin{eqnarray}
\label{e26a}
\la x^2\ra_{\tau}^\mathrm{eq}&=&\frac{\hbar}{2M\omega_1}\,\coth{\left(\beta\hbar\omega_1/2 \right)}\\
\label{e26b}
\la p^2\ra_{\tau}^\mathrm{eq}&=&\frac{\hbar \omega_1 M}{2}\,\coth{\left(\beta\hbar\omega_1/2 \right)}\,.
\end{eqnarray}
\end{subequations}
On the other hand,  for the final nonequilibrium state $\rho_\tau$, we have  $\la x\ra_{\tau}=\la p\ra_{\tau}=0$, and 
\begin{subequations}
\begin{eqnarray}
\label{e27a}
\la x^2\ra_{\tau}&=&\frac{\hbar}{2M\omega_0}\,\left(Y_\tau^2+\omega_0^2 X_\tau^2 \right)\,\coth{\left(\beta\hbar\omega_0/2 \right)}\\
\label{e27b}
\la p^2\ra_{\tau}&=&\frac{\hbar  M}{2\omega_0}\,\left(\dot Y_\tau^2+\omega_0^2 \dot X_\tau^2 \right)\,\coth{\left(\beta\hbar\omega_0 /2\right)}.
\end{eqnarray}
\end{subequations}
The cross correlation function  can be evaluated by 
 exploiting the fact that $\la xp+px \ra_\tau=M\, \td_t\la x^2\ra_{\tau}$, and reads 
 \begin{equation}
\label{e28}
\la xp+px \ra_\tau=\frac{\hbar}{\omega_0}\,\left(Y_\tau \dot Y_\tau+\omega_0^2 X_\tau \dot X_\tau \right)\,\coth{\left(\beta\hbar\omega_0/2 \right)}\,.
\end{equation}
The analytic expression of the quantum fidelity function $F(\rho_\tau,\rho_\tau^\mathrm{eq})$ between  nonequilibrium and equilibrium oscillator states  at the end of the driving can be finally obtained by evaluating the determinants $\Delta$ and $\delta$ in Eq.~\eqref{e23} using Eqs.~\eqref{e26a}-\eqref{e28}, with the help of  the relation $\dot X_t Y_t-X_t\dot Y_t=1$ \cite{hus53} and the definition of the function $Q^*$ given in Eq.~\eqref{e21}. We find, 
\begin{widetext}
\begin{equation}
\label{e29}
F\left(\rho_\tau,\rho_\tau^\mathrm{eq}\right)=\frac{2}{\sqrt{\mathrm{ct}^2{\left(\beta\varepsilon_0 /2\right)}+\mathrm{ct}^2{\left(\beta\varepsilon_1/2 \right)}+2 Q^{*}\,\mathrm{ct}{\left(\beta\varepsilon_0 /2\right)}\mathrm{ct}{\left(\beta\varepsilon_1/2 \right)}+\mathrm{c}^2{\left(\beta\varepsilon_0 /2\right)}\mathrm{c}^2{\left(\beta\varepsilon_1/2 \right)}}-\mathrm{c}{\left(\beta\varepsilon_0/2 \right)}\mathrm{c}{\left(\beta\varepsilon_1/2 \right)}},
\end{equation}
\end{widetext}
with the notation $\mathrm{c}(.)=\mathrm{csch}(.)$ and $\mathrm{ct}(.)=\mathrm{coth}(.)$, and  the energies $\varepsilon_i=\hbar\omega_i$. The Bures angle $\mathcal{L}$ then directly follows from Eq.~\eqref{e05}, and the lower bound to the nonequilibrium entropy production $\la \Sigma \ra$ can be determined, to any order, with the help of  the expansion \eqref{e09}. 

To get more physical insight, let us evaluate the limiting expressions of the fidelity \eqref{e29} in the low-temperature (quantum) and high-temperature (classical) regimes. An expansion of the hyperbolic cosine and cotangent functions in the zero-temperature limit, $\hbar\beta\rightarrow\infty$, leads to,
\begin{equation}
\label{e30}
F\left(\rho_\tau,\rho_\tau^\mathrm{eq}\right)\stackrel{\hbar\beta\rightarrow \infty}{\longrightarrow}\sqrt{\frac{2}{1+Q^*}}\,.
\end{equation}
In the adiabatic limit $Q^*\rightarrow 1$, the fidelity thus tends to one, that is the Bures angle approaches zero,  indicating that the system ends in an equilibrium state, as expected. For strongly nonadiabatic processes, $Q^*\gg1$, on the other hand, the fidelity tends to zero as $1/\sqrt{Q^*}$. Here the Bures angle tends to $\pi/2$, showing that $\rho_\tau$ and   $\rho_\tau^\text{eq}$ are maximally distinguishable (orthogonal).

 Equation \eqref{e30} can also be derived directly by noting that   in the zero-temperature limit, the harmonic oscillator is initially in a pure state $\ket{0_0}$. The initial, equilibrium density operator is, hence, $\rho_0|_{T=0}=\ket{0_0}\bra{0_0}$ and analogously for $\rho_\tau^\mathrm{eq}\big|_{T=0}$. Since these states are pure,  the fidelity simplifies to their overlap, and we have, $F\left( \rho_\tau,\rho_\tau^\mathrm{eq}\right)|_{T=0}=\tr{\rho_\tau\,\rho_\tau^\mathrm{eq}}=p_{0,0}^\tau$, where $p_{0,0}^\tau$ is the probability for the system to start and end in the corresponding ground state. The latter  is given by the expression \cite{def10},
\begin{equation}
\label{e31}
p_{0,0}^\tau=\sqrt{\frac{2}{1+Q^*}},
\end{equation}
and we thus recover  Eq.~\eqref{e30}.

In the classical limit, $\hbar\beta\rightarrow0$, by repeating the same analysis, the fidelity \eqref{e29} simplifies to,
\begin{equation}
\label{e32}
F\left( \rho_\tau,\rho_\tau^\mathrm{eq}\right)\stackrel{\hbar\beta\rightarrow 0}{\longrightarrow}\frac{4\,\omega_0\omega_1}{\omega_0^2+2\, Q^* \omega_0\omega_1+\omega_1^2}\,.
\end{equation}
For an adiabatic frequency change, $Q^*\rightarrow 1$, the fidelity  reduces to $F\left( \rho_\tau,\rho_\tau^\mathrm{eq}\right) \simeq 4 \omega_0\omega_1/(\omega_0+\omega_1)^2$. Therefore, as noticed in Ref.~\cite{def08} (see also Ref.~\cite{all05}), a unitary process can only be quasistatic in the thermodynamic sense, $F\left( \rho_\tau,\rho_\tau^\mathrm{eq}\right) \simeq 1$, if $|\omega_1- \omega_0|/\omega_0\ll 1$. 
In addition, we note that in the classical limit the fidelity vanishes for large values of $Q^*$ as $1/Q^*$, that is much faster than in the low-temperature regime. The density operators $\rho_\tau$ and   $\rho_\tau^\text{eq}$ thus become orthogonal ($\mathcal{L}=\pi/2$) much faster as a function of the degree of nonadiabaticity.

For the parametric quantum oscillator, the nonequilibrium entropy production \eqref{e07} can be determined exactly, allowing to test the generalized Clausius inequality \eqref{e10}. It is given by \cite{def08},
\begin{equation}
\label{e34}
\begin{split}
\la \Sigma\ra=&\frac{\beta}{2}\,\left(Q^* \hbar\omega_1-\hbar\omega_0\right)\coth\left(\beta\hbar\omega_0/2\right)\\
-&\ln\left(\frac{\sinh\left(\beta\hbar\omega_1/2\right)}{\sinh\left(\beta\hbar\omega_0/2\right)}\right),
\end{split}
\end{equation}
where we used $\la H_0\ra=(\hbar\omega_0/2)\coth(\beta \hbar\omega_0/2)$ and Eq.~\eqref{e22}. Figure~1 shows the nonequilibrium  entropy production  $\la \Sigma\ra$ as a function of the measure of adiabaticity $Q^*$, together with the lower bound obtained with the first term in the expansion \eqref{e09} and  the exact function $s(x)$ \eqref{e08} (the latter is indistinguishable from the expression including the first five nonzero terms of the expansion). We see that the first term in the expansion provides a good lower bound in many cases.

\begin{figure}[h]
\centering
\includegraphics[width=0.47\textwidth]{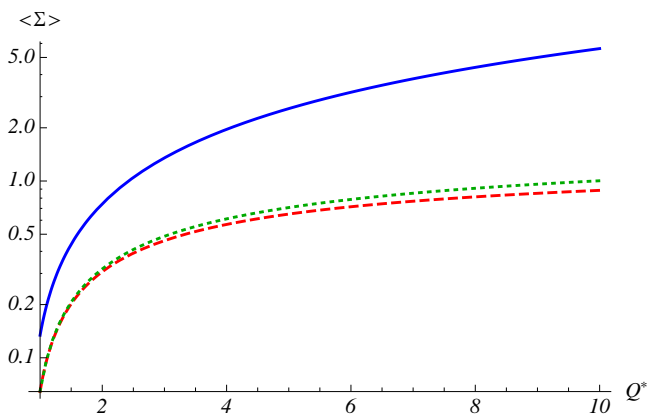}
\caption{(Color online) Irreversible entropy production $\la \Sigma\ra$ \eqref{e34} (blue, solid line) together with the lower bound \eqref{e10} corresponding to the lowest order term of the expansion \eqref{e09}(red, dashed line), or the exact function $s(x)$ \eqref{e08} (green, dotted line), as a function of the adiabaticity parameter $Q^*$ \eqref{e21}. Parameters are  $\hbar=1$, $\beta=1.2$, $\omega_0=0.9$, $\omega_1=0.5 $.} 
\label{claus_ineq}
\end{figure}

\section{Lower bound based on the trace distance}
\begin{figure}[h]
\centering
\includegraphics[width=0.47\textwidth]{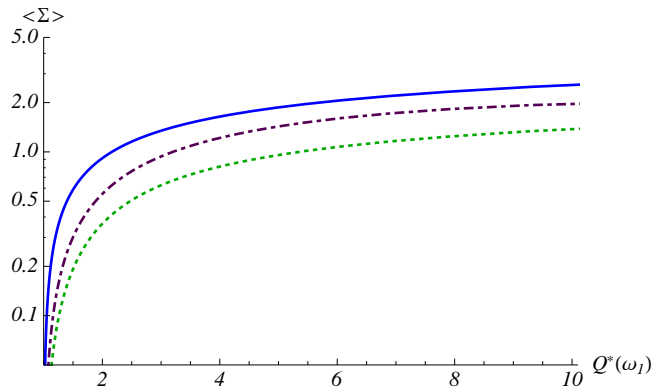}
\caption{(Color online) Irreversible entropy production $\la \Sigma\ra$ \eqref{e34} (blue, solid line) together with the lower bound \eqref{e10} evaluated for the Bures angle $\mc{L}$ \eqref{e05} (green, dotted line), and the trace distance $\mc{T}$ \eqref{e34a} (purple, dotdashed line), as a function of the adiabaticity parameter  \eqref{e21} for the sudden switch $Q^*(\omega_1)= (\omega_0^2+\omega_1^2)/(2\omega_0\omega_1)$. Parameters are  $\hbar=1$, $\beta=4.8$, $\omega_0=0.9$.} 
\label{claus_ineq_2}
\end{figure}

As discussed in Section I, the Bures angle $\mathcal{L}$, being the extension of Wootters' statistical distance to mixed states,  possesses a simple interpretation as the geometric angle between two density operators. However, Eq.~\eqref{e07} shows that the nonequilibrium entropy production $\la \Sigma\ra$ is bounded by many unitarily invariant distances, albeit with possibly less natural physical interpretation. To elucidate this point, we discuss the concrete case of the trace norm, which has been reported to yield the largest lower bound on the relative entropy \cite{aud05} (a further comparison of $\mc{L}$ with the Bures distance $\mathcal{D}$ is given in Appendix A). The trace distance between two density operators, $\rho_1$ and $\rho_2$, is defined as \cite{nie00,ben06}
\begin{equation}
\label{e34a}
 \mc{T}(\rho_1,\rho_2) = \frac{1}{2} \tr{|\rho_1-\rho_2|} = \frac{1}{2} \tr{\sqrt{(\rho_1-\rho_2)^2}}
\end{equation}
Contrary to the Bures angle (or the Bures distance), it is not a Riemannian distance---however, both are monotone. The trace distance between nonequilibrium and equilibrium states of the parametric quantum oscillator \eqref{e18} can be evaluated for arbitrary driving with the help of the explicit expressions of $\rho_\tau$ and $\rho_\tau^\text{eq}$ given in Appendix C: we have $\mc{T}(\rho_\tau,\rho_\tau^\text{eq}) = (1/2) \sum_i |\lambda_i|$, where $\lambda_i$ are the eigenvalues of $\rho_\tau-\rho_\tau^\text{eq}$. Unlike for the Bures angle, it does not seem  to be possible to express $\mc{T}$ as a function of the adiabaticity parameter $Q^*$ alone (the density operators $\rho_\tau$ and $\rho_\tau^\text{eq}$ depend on the two functions $X_t$ and $Y_t$, and not on $Q^*$ directly). To circumvent this problem, we have numerically evaluated the trace distance for the case of a sudden switching of the frequencies for which $Q^* = (\omega_0^2+\omega_1^2)/(2\omega_0\omega_1)$. 
Figure \ref{claus_ineq_2} shows the corresponding  entropy production $\la \Sigma\ra$ \eqref{e34} and the lower bound \eqref{e10} for the Bures angle, $s(2\mathcal{L}/\pi)$, and for the trace distance,  $s(\mc{T})$,  as a function of $Q^*$ for fixed $\beta$ and $\omega_0$. Contrary to Fig.~\ref{claus_ineq}, where both $\omega_0$ and $\omega_1$ are fixed, for the sudden frequency switch $Q^*(\omega_1)$ is a function of $\omega_1$.

We observe that the lower bound based on the trace distance is sharper  than the one obtained using the Bures angle. However, the two bounds appear  largely equivalent, reflecting the fact that  $\mathcal{L}$ and $\mc{T}$ are closely related (see e.g.  \cite{nie00,ben06}). We stress that the trace distance lacks the simple interpretation of the Bures angle as the  angle between density operators. Moreover, in the classical limit the bound based on the trace distance does not reduce to the known bound on the entropy production derived in linear response theory \cite{sal83,sal84,nul85}.

\section{Conclusions}
The Bures angle between the nonequilibrium and the corresponding equilibrium state of a driven closed quantum system defines a thermodynamic length that is valid arbitrarily far from equilibrium. The latter can be used to characterize the departure from equilibrium for generic driving. We derived a lower bound on the nonequilibrium entropy production, which we expressed as a function of the Bures angle,  by using a sharp lower bound on the quantum relative entropy. In such a way, we obtained a generalized Clausius inequality, with a process-dependent lower bound, that holds beyond the range of linear response theory. As an illustration, we treated the case of a time-dependent harmonic oscillator for which we derived analytic expressions for the Bures angle. We further compared the lower bound obtained with the Bures angle with the one based on the trace distance. While the trace distance offers a slightly sharper bound, the two appear to be largely equivalent.

\acknowledgments{This work was supported by  the DFG (contract No LU1382/4-1). SD  acknowledges financial support by a fellowship within the postdoc-program of the German Academic Exchange Service (DAAD, contract No D/11/40955).}

\appendix
 \section{Lower bound based on the Bures distance}
\begin{figure}
\centering
\includegraphics[width=0.47\textwidth]{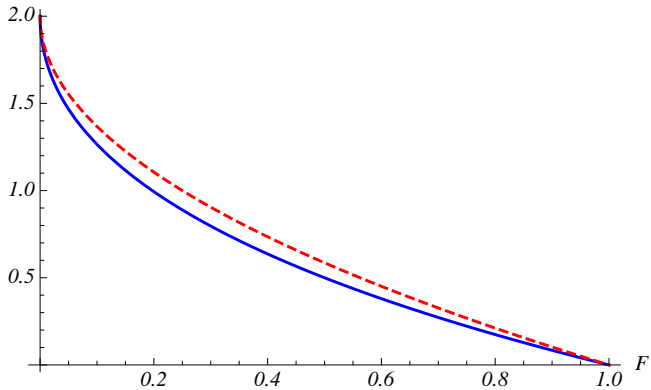}
\caption{(Color online) Lower bounds for the nonequilibrium entropy production based on the Bure angle,  $(8/\pi^2) \mathcal{L}^2$ \eqref{e10} (blue  solid), and on the Bures distance, $\mathcal{D}^2$ \eqref{e17} (red dashed), as a function of the fidelity $F$.}
\label{fig1}
\end{figure}

To evaluate the changes induced by the choice of another unitarily invariant distance on the lower bound \eqref{e10}, we present in this Appendix an alternative, constructive derivation of the lowest order estimation of the nonequilibrium entropy production  $\la \Sigma\ra$. 
Let us begin by  introducing the Hellinger distance \cite{hel09},
\begin{equation}
\label{e11}
\mathcal{H}^2 \left( p_1,p_2\right) =\int \td x\,\left( \sqrt{p_1\left( x\right) }-\sqrt{p_2\left( x\right) } \right) ^2\,,
\end{equation}
for two (classical) probability distributions $p_1(x)$ and $p_2(x)$. The Hellinger distance is  another measure of the distinguishability of two probability distributions. It is a true distance which fulfills  symmetry, non-negativity and the triangle inequality. Expression  \eqref{e11} can be rewritten in terms of the classical fidelity function, $f(p_1,p_2)=\int\td x \sqrt{p_1(x)p_2(x)}$, to yield
\begin{equation}
\label{e12}
\mathcal{H} \left( p_1,p_2\right) =\sqrt{2- 2\, f\left( p_1,p_2\right) }.
\end{equation}
By using  the inequality, $\sqrt{y}-1\geq 1/2\,\ln{\left( y\right) }$, we have
\begin{equation}
\label{e13}
\sqrt{\frac{p_2\left( x\right) }{p_1\left( x\right) }}-1\geq\frac{1}{2}\left(\ln{p_2(x)}-\ln{p_1(x)}\right).
\end{equation}
Averaging  Eq.~\eqref{e13} over the probability distribution $p_1(x)$ results in,
\begin{equation}
\label{e14}
2\left(1-\la \sqrt{\frac{p_2\left( x\right) }{p_1\left( x\right) }}\ra_{p_1}\right) \leq\la \ln{p_1(x)}-\ln{p_2(x)}\ra_{p_1},
\end{equation}
from which we deduce the inequality,
\begin{equation}
\label{e15}
D\left( p_1||p_2\right) \geq 2- 2\,f\left( p_1,p_2\right) = \mathcal{H}^2 \left( p_1,p_2\right).
\end{equation}
Here $D( p_1||p_2)$ denotes the classical Kullback-Leibler divergence between $p_1$ and $p_2$.  The  classical result  \eqref{e15} can be extended  to  quantum states by considering the quantum version of the Hellinger distance which is the Bures distance  between density operators  $\rho_1$ and $\rho_2$ \cite{joz94},
\begin{equation}
\label{e16}
\mathcal{D}^2\left(\rho_1,\rho_2\right)=2\left(1-\sqrt{F\left(\rho_1,\rho_2\right)}\right)\,.
\end{equation}
Note the difference in the definitions of the classical and quantum fidelity. By combining Eqs.~\eqref{e04}, \eqref{e15} and \eqref{e16}, we then find the generalized Clausius inequality,
\begin{equation}
\label{e17}
\la \Sigma\ra \geq \mathcal{D}^2\left(\rho_\tau,\rho_\tau^\mathrm{eq}\right)\,.
\end{equation}
The above lower bound on the entropy production corresponds to the lowest order term in the expansion  \eqref{e09} when the Bures distances is chosen instead of the Bures angle in Eq.~\eqref{e07}, since $\mathcal{D}(e^{1,1},e^{2,2})=\sqrt{2}$. Figure~\ref{fig1} shows the Bures angle  $\mathcal{L}$ and the Bures distance $\mathcal{D}$ as a function of the quantum fidelity $F$. We observe that $\mathcal{D}\geq \mathcal{L}$ so that the Bures distance offers a (slightly) sharper bound to the entropy production than the Bures angle, to lowest order. However, the distance $\mathcal{D}$ bears the disadvantage that the intuitive, physical interpretation as an angle between mixed states is lacking.

\section{Analytic upper bound for the quantum relative entropy}

Due to the importance of the relative entropy in physics and mathematics, and the complexity to evaluate it, accurate approximations and bounds are essential. While a lower bound has been obtained for unitarily invariant norms in the form of Eq.~\eqref{e07} \cite{aud05}, upper bounds are much more difficult to find. Recently, a general upper bound was proposed in terms of the eigenvalues of the density operators \cite{aud11}. In the present thermodynamic context, we may, however,  derive a simpler upper bound.
We start with the inequality \cite{heb01},
\begin{equation}
\label{e35}
\tr{\rho_1 \ln{\rho_1}-\rho_1\ln{\rho_2}}\leq \frac{1}{\nu} \tr{\rho_1^{1+\nu}\,\rho_2^{-\nu}-\rho_1}\,,
\end{equation}
which is true for all positive definite operators $\rho_1, \rho_2$ and $\nu>0$. We shall here  concentrate on the final nonequilibrium and equilibrium density operators, $\rho_\tau$ and $\rho_\tau^\mathrm{eq}$. By  choosing $\nu=1$ and using the normalization condition $\tr{\rho}=1$, we obtain the upper bound, $\la \Sigma\ra \leq \tr{\rho_\tau^2\,\left( {\rho_\tau^\mathrm{eq}}\right)^{-1}}-1$. 
In order to further simplify the bound and derive an expression which does not depend on the off diagonal matrix elements in energy representation of the density operators, we use the inequality \cite{mir75},
\begin{equation}
\label{e36}
\left|\tr{\rho_1 \rho_2} \right|\leq \sum\limits_{r=1}^n \sigma^1_r \sigma_r^2,
\end{equation}
which holds for any complex $n\times n$ matrices $\rho_1$ and $\rho_2$ with descending singular values, $\sigma^1_1\geq...\geq\sigma^1_n$ and $\sigma^2_1\geq...\geq\sigma^2_n$. The singular values of an operator $A$ acting on a Hilbert space are defined as the eigenvalues of the operator $\sqrt{A^\dagger A}$. If  $\rho_1$ and $\rho_2$ are density operators acting on the same Hilbert space Eq.~\eqref{e36} remains true for arbitrary dimensions, and the singular values are identically given by the eigenvalues \cite{gri91}. As a result, we obtain the  upper bound for the entropy production $\la \Sigma\ra$,
\begin{equation}
\label{e37}
\la \Sigma\ra\leq \sum\limits_n \frac{\left(p_n^\tau\right)^2}{p_n^\mathrm{eq}}-1.
\end{equation}
%

\section{Explicit expressions of the density operators \label{app2}}

The evaluation of the covariance matrix \eqref{e24} requires the expressions of the density operators $\rho_0^\text{eq}$, $\rho_\tau^\text{eq}$  and $\rho_\tau$ in coordinate representation. We collect them in this Appendix for completeness.  The initial equilibrium density operator  $\rho_0^\text{eq}$ is given by \cite{gre87}, 
\begin{equation}
\label{e40}
\begin{split}
&\rho_0^\text{eq}(x,y)=\sqrt{\frac{M\omega_0}{\pi\hbar}\tanh{\left(\beta\hbar\omega_0/2 \right)}}\times\\
&\exp{\left(-\frac{M\omega_0}{2\hbar}\coth{\left(\beta\hbar\omega_0\right)}\left(x^2+y^2 -2\,\mathrm{sech}{\left(\beta\hbar\omega_0 \right)} x y \right)\right)}\,.
\end{split}
\end{equation}
The final equilibrium density operator $\rho_\tau^\text{eq}$ has the same form as Eq.~\eqref{e40} with the replacement $\omega_0$ by $\omega_1$.
On the other hand, the final nonequilibrium  operator $\rho_\tau$ can be derived  from Eq.~\eqref{e40} by noting that $\rho_\tau(x,x')=\int\td y\int\td y' \, U_\tau(x,y)\,\rho_0^\text{eq}(y,y')\,U_\tau^*(y',x')$. The propagator $U_\tau(x,x_0)$ can be determined from the wave function \eqref{e19} with $\psi_\tau(x) = \int \td x_0 \,U_\tau(x,x_0)\, \psi_{t_0}(x_0)$, and reads \cite{hus53},
\begin{equation}
\label{e42}
U_\tau= \sqrt{\frac{M}{2\pi i \hbar X_\tau}} \exp{\left(\frac{i M}{2\hbar X_\tau} \left[\dot X_\tau x^2-2 xx_0+ Y_\tau x_0^2\right] \right)}
\end{equation}
The functions $X_\tau$ and $Y_\tau$ are  solutions of the force free harmonic  oscillator satisfying the boundary conditions $X_0=0$, $\dot X_0=1$ and $Y_0=1$, $\dot Y_0=0$. A direct evaluation of the Gaussian integrals leads to the expression,
\begin{widetext}
\begin{equation}
\label{e43}
\begin{split}
&\rho_\tau(x,y)=\sqrt{\frac{M\omega_0}{\pi\hbar}\frac{\tanh{\left(\beta\hbar\omega_0/2 \right)}}{Y_\tau^2+\omega_0^2 X_\tau^2}}\times
\\ &  \exp{\left(-\frac{M\omega_0}{2\hbar}\frac{1}{Y_\tau^2+\omega_0^2 X_\tau^2}\left[\coth{\left(\beta\hbar\omega_0\right)}\left(x^2+y^2-2\,\mathrm{sech}{\left(\beta\hbar\omega_0 \right)} x y \right)+ i \left(x^2-y^2\right)\left(\omega_0^2 \dot X_\tau X_\tau+\dot Y_\tau Y_\tau\right)\right]\right)}.
\end{split}
\end{equation}
\end{widetext}


\begin{thebibliography}{99}
\bibitem{rem1} All the coefficients in the expansion are not positive; the first negative one is the coefficient of $x^{62}$ \cite{aud05}.
\end{thebibliography}
\end{document}